\documentclass[twocolumn,aps,prd,showpacs,nofootinbib,superscriptaddress,
preprintnumbers,floatfix,10pt]{revtex4-1}

\usepackage{amsmath}
\usepackage{amsfonts}
\usepackage{amssymb}
\usepackage{epsfig}
\usepackage{graphicx}
\usepackage{color}
\usepackage{slashed}
\usepackage{epstopdf}
\usepackage{dsfont}

\usepackage{hyperref}

\newcommand{\Eqref}[1]{Eq.~\eqref{#1}}

\newcommand{\nablaTild}{\tilde{\nabla}}
\newcommand{\GammaTild}{\tilde{\Gamma}}

\newcommand{\RTild}{\tilde{R}}
\newcommand{\LTild}{\tilde{L}}

\newcommand{\sigmaBar}{\bar{\sigma}}

\newcommand{\gBar}{\bar{g}}
\newcommand{\ABar}{\bar{A}}

\newcommand{\nablaBar}{\bar{\nabla}}

\newcommand{\PhiBar}{\bar{\Phi}}

\begin{document}

\preprint{}

\title{Asymptotically Safe Hilbert-Palatini Gravity in an On-Shell Reduction 
Scheme} 

\author{Holger Gies}
\email{holger.gies@uni-jena.de}
\affiliation{Theoretisch-Physikalisches Institut, 
Abbe Center of Photonics, Friedrich Schiller University Jena, Max Wien 
Platz 1, 07743 Jena, Germany}
\affiliation{Helmholtz-Institut Jena, Fr\"obelstieg 3, D-07743 Jena, Germany}
\affiliation{GSI Helmholtzzentrum für Schwerionenforschung, Planckstr. 1, 
64291 Darmstadt, Germany} 
\author{Abdol Sabor Salek}
\email{abdol.sabor.salek@uni-jena.de}
\affiliation{Theoretisch-Physikalisches Institut, 
Abbe Center of Photonics, Friedrich Schiller University Jena, Max Wien 
Platz 1, 07743 Jena, Germany}

\begin{abstract}
We study the renormalization flow of Hilbert-Palatini gravity to lowest 
non-trivial order. We find evidence for an asymptotically safe high-energy 
completion based on the existence of an ultraviolet fixed point similar to the 
Reuter fixed point of quantum Einstein gravity. 

In order to manage the quantization of the large number of independent degrees 
of freedom in terms of the metric as well as the connection, we use an on-shell 
reduction scheme: for this, we quantize all degrees of freedom beyond 
Einstein gravity at a given order that remain after using the equations of 
motion at the preceding order. In this way, we can straightforwardly keep track 
of the differences emerging from quantizing Hilbert-Palatini gravity in 
comparison with Einstein gravity. To lowest non-trivial order, the difference is 
parametrized by fluctuations of an additional abelian gauge field.

The critical properties of the ultraviolet fixed point of Hilbert-Palatini 
gravity are similar to those of the Reuter fixed point, occurring at a smaller 
Newton coupling and exhibiting more stable higher order exponents. 

%
%
\end{abstract}


\maketitle

\section{Introduction}
\label{intro}

For the approach to quantizing gravity both the degrees of freedom to be 
quantized as well as the correct quantization method are a matter of intense 
debate. A priori different choices could lead to different potentially 
consistent theories of quantum gravity ultimately requiring experimental data 
to single out the theory realized in nature. 

It is well known that -- already on the classical level -- very different 
choices for the degrees of freedom can lead to the same classical equation of 
motion, namely Einstein's equation 
\cite{Kibble:1961ba,Sciama:1964wt,Krasnov:2017epi,Krasnov:2020lku}. This 
includes of course the maybe simplest choice being the metric, but also the 
vierbein, possibly in combinations with various forms of connections. Of 
course, classical equivalence does not entail quantum equivalence, therefore 
different choices might rather be expected to lead to different quantum 
theories. 

In turn, different choices of degrees of freedom or even the quantization 
procedure could finally describe the same quantum theory if they lead to the 
same universality class as for instance identified by a renormalization group 
analysis. In discrete approaches this can be indicated by the presence of a 
second order phase transition or even quantified in terms of critical exponents 
of a corresponding quantum critical point; in the context of gravity, cf. 
\cite{Ambjorn:2012jv,Coumbe:2014nea,Bahr:2016hwc,Asaduzzaman:2022kxz}. 

Research in recent years has accumulated evidence that such a quantum 
critical point exists for gravity when using the metric as the fundamental 
degree of freedom together with a (standard) quantization procedure that is 
able to capture non-perturbative information. The latter is necessary, since 
the quantum critical point corresponds to a fixed point of the renormalization 
group at a finite coupling, realizing Weinberg's asymptotic safety scenario for 
Einstein gravity \cite{Weinberg:1976xy,Weinberg:1980gg}. This Reuter fixed 
point has been discovered by applying functional renormalization group 
(RG) methods to gravity \cite{Reuter:1996cp}, and confirmed in many refined 
studies, see 
\cite{Percacci:2017fkn,Reuter:2019byg,Pereira:2019dbn,Pawlowski:2020qer,
Bonanno:2020bil,Reichert:2020mja} for recent reviews.

Functional RG methods and a classification of universality classes 
\cite{Martini:2021lcx} are not limited to the metric as the quantum degree 
of freedom. In fact, pioneering studies have been performed for Einstein-Cartan 
theory with the Hilbert-Palatini action being generalized to the Holst action 
\cite{Daum:2010qt,Daum:2013fu,Harst:2014vca} or constrained to self-dual 
connections \cite{Harst:2015eha}, or for ``tetrad only'' formulations 
\cite{Harst:2012ni}. All these works find indications for the existence of UV 
fixed points supporting asymptotic safety of such quantum gravity theories, 
with the fixed points likely representing universality classes different from 
that of quantum Einstein gravity. In fact rather complex phase diagrams are 
partly found being paralleled by the complexity of the computations involving 
such a large number of gauge degrees of freedom. On the other hand, it is 
interesting to observe that also reduced versions such as unimodular gravity 
\cite{Eichhorn:2013xr,Eichhorn:2015bna,deBrito:2019umw,deBrito:2021pmw,
deBrito:2020xhy} or conformally reduced gravity 
\cite{Reuter:2008wj,Machado:2009ph,Dietz:2015owa,Nagy:2019jef} exhibit UV 
complete renormalization group trajectories (see however \cite{Knorr:2020ckv} 
for a critical view on conformally reduced gravity.)   

A major motivation to study formulations of gravity based on the metric 
and the 
connection is the greater similarity to gauge theories of particle physics. In 
addition to this structural resemblance, also a larger technical toolkit 
developed for gauge theories may become available for concrete 
calculations; 
specifically lattice formulations for gravity become accessible 
\cite{Schaden:2015wia,Asaduzzaman:2019mtx}. 

In the present work, we suggest to reduce the amount of complexity 
introduced 
by the large number of gauge degrees of freedom of a metric-affine 
formulation 
by an on-shell reduction scheme: at a given expansion order, we only 
quantize 
those degrees of freedom which remain after using the equations of 
motion of 
the preceeding order. For instance at lowest order in the curvature 
(Einstein-Hilbert level), different choices of degrees of freedom boil down to 
the Einstein equation for the metric; hence the on-shell reduction 
suggests to 
quantize only the metric at this level. At higher order in the curvature, 
connection degrees of freedom typically develop their independent 
dynamics. In 
the present work, we will focus on the second-order curvature level 
where a 
co-vector field remains as an independent degree of freedom in the 
connection after on-shell reduction. The corresponding additional 
action to be quantized is of Maxwell type. 

In the asymptotic safety scenario, this on-shell reduction scheme helps 
to monitor the quantitative modifications of the RG flow and of the 
universality 
class associated with RG fixed points in a controlled and systematic way. 
We observe a UV-attractive fixed point of Reuter type with more 
stabilized critical exponents at a smaller value of the Newton coupling $G$.

The present paper is organized as follows: in Sect.~\ref{sec:EPGrav}, we 
review Hilbert-Palatini gravity to second order in the curvature and apply 
on-shell reduction in order to identify the degrees of freedom to be 
quantized. The correspondingly quantized theory is studied in 
Sect.~\ref{sec:QEPG} using the functional RG. Here, we focus on the UV 
fixed-point structure and determine the critical parameters in comparison to 
those of metric gravity. We conclude in Sect.~\ref{sec:conc}.

\section{Hilbert-Palatini gravity}
\label{sec:EPGrav}

In the Einstein formulation of gravity, the connection is linked to the metric 
in the form of the Levi-Civita connection. By contrast, the Hilbert-Palatini 
formulation of gravity treats the metric and the 
connection as independent degrees of freedom. Therefore, the connection 
can a priori carry additional degrees of freedom that may or may not be fully 
linked to the metric via the equations of motion. In the following, we start 
with a general connection and study the on-shell constraints at increasing 
orders of curvature. 

\subsection{General connection on smooth manifolds}
\label{subsec:GenCon}

Suppose we have a vector field $V$ on a smooth manifold with 
metric $g$ and connection $\GammaTild$. The covariant 
derivative $\nablaTild$ of this vector field reads
\begin{eqnarray}
\nablaTild_\mu V^\alpha = \partial_\mu V^\alpha + 
\GammaTild^\alpha_{\mu \nu} V^\nu .
\end{eqnarray}
In the most general case, the connection $\GammaTild$ can be 
decomposed into the expressions for the Levi-Civita connection $\Gamma$ 
(Christoffel symbols), contorsion $K$ and displacement $L$ 
(though the latter does not have a collectively agreed upon name
\cite{Baldazzi:2021kaf},
\begin{eqnarray}
\GammaTild^\alpha_{\mu \nu} = \Gamma^\alpha_{\mu \nu} + 
K^\alpha_{\mu \nu} + L^\alpha_{\mu \nu}.
\label{eq:genconnec}
\end{eqnarray} 
The Levi-Civita connection is constructed from the metric, 
$\Gamma=\Gamma[g]$ and can be used to define the standard covariant derivative 
$\nabla$ of Einstein gravity
\begin{eqnarray}
\nabla_\mu V^\alpha = \partial_\mu V^\alpha + \Gamma^
\alpha_{\mu \nu} V^\nu .
\end{eqnarray}
The Levi-Civita part of the connection $\Gamma$ accounts for 
curvature through the Riemann tensor defined below. The contorsion 
tensor $K$ is related to Cartan torsion $T$ which corresponds to 
the anti-symmetric part of the connection with respect to the lower 
indices,
\begin{eqnarray}
T^\alpha_{\mu \nu} &= \GammaTild^\alpha_{\mu \nu} - 
\GammaTild^\alpha_{\nu \mu}.
\end{eqnarray}
The displacement tensor $L$ induces non-metricity $Q$ which is 
symmetric in the last two indices,
\begin{eqnarray}
\nablaTild_\mu g_{\alpha \beta}= - Q_{\mu \alpha \beta}.
\end{eqnarray}
the linear relation between contorsion $K$ and torsion $T$, and 
analogously for the displacement $L$ and the non-metricity $Q$, can be found in 
the literature \cite{Baldazzi:2021kaf,Vitagliano:2013rna,Hehl:1994ue}, but are 
not of relevance for what follows.

The general connection \eqref{eq:genconnec} can be used to define a generalized 
Riemann tensor analogously to the pure metric formulation,
\begin{eqnarray}
\left[ \nablaTild_\mu , \nablaTild_\nu \right] V_\sigma =
\RTild^\rho{}_{\sigma \mu \nu} V_\rho ,
\end{eqnarray}
taking the familiar form
\begin{eqnarray}
\RTild^\rho{}_{\sigma \mu \nu} &=& \partial_\mu \GammaTild^
\rho_{\nu \sigma} - \partial_\nu \GammaTild^\rho_{\mu 
\sigma} + \GammaTild^\rho_{\mu \lambda} \GammaTild^
\lambda_{\nu \sigma} - \GammaTild^\rho_{\nu \lambda} 
\GammaTild^\lambda_{\mu \sigma}. \label{eq:Rtilde}
\end{eqnarray}
In contrast to the standard case, where the Riemann tensor is composed out of 
the metric, we can view $\RTild$ as dependent on the metric, the contorsion and 
the displacement, $\RTild=\RTild[g,K,L]$. 
Since the general connection $\GammaTild$ is not symmetric 
in the lower two indices anymore, the generalized Riemann tensor $
\RTild^{\cdot}{}_{\cdot\cdot\cdot}$ is not anti-symmetric in the first two 
indices, as will be important below.

Analogously, we can construct a generalized Ricci tensor by contracting the 
first and the third index
\begin{eqnarray}
\RTild_{\sigma \nu} &=& \RTild^\rho{}_{\sigma \mu \nu} \ g^
\mu_\rho ,
\end{eqnarray}
which - contrary to Einstein gravity - is not purely symmetric anymore. 
Contracting this tensor further 
leads to a generalized Ricci scalar, 
\begin{eqnarray}
\RTild &=& \RTild_{\sigma \nu} g^{\sigma \nu} .
\end{eqnarray}
As in Einstein gravity, we can use these generalized curvature forms 
to construct curvature invariants and formulate an action $S$ governing the 
dynamics of a theory, with the metric, the contorsion/torsion and the 
displacement/non-metricity as fundamental degrees of freedom, $S=S[g,K,L]$. 
In a quantized version, all these degrees of freedom have to be integrated out, 
requiring appropriate gauge fixing also for the connection degrees of freedom, 
cf. for instance \cite{Daum:2010qt}. 

\subsection{Einstein-Hilbert-Palatini action}
\label{subsec:PalatiniEinsteinHilbert}

Let us focus on a Palatini formulation of gravity starting with the lowest 
nontrivial order in the curvature. This corresponds to the 
Einstein-Hilbert action (also referred to as 
Einstein-Hilbert-Palatini action in order to emphasize the dependence on the 
general connection),
\begin{eqnarray}
S[g,\GammaTild] = \int d^4x \sqrt{g} \frac{1}{16 \pi G} \left( 
\Lambda - 2 \RTild \right).
\label{eq:EHP}
\end{eqnarray}
Classically, the corresponding fields (the metric $g$ and the 
connection $\GammaTild$) are constrained by their equations 
of motions, namely
\begin{eqnarray}
\label{eq:PalatiniEqMetric}
\frac{\delta S}{\delta g_{\mu \nu}} &\overset{!}{=}& 0
\end{eqnarray}
which is a partial differential equation for the metric, 
and additionally in the Palatini formulation
\begin{eqnarray}
\label{eq:PalatiniEqConnection}
\frac{\delta S}{\delta \GammaTild^\alpha_{\mu \nu}} &
\overset{!}{=}& 0 .
\end{eqnarray}
This new equation can be interpreted as an equation of motion for the 
contorsion $K$ and displacement $L$. As the derivative terms turn out to be 
total derivatives, \Eqref{eq:PalatiniEqConnection} is a purely algebraic 
equation for $K$ and $L$. Its general solution in terms of the general 
connection $\GammaTild$ can be written as
\begin{eqnarray}
\label{eq:genPalatiniSol}
\GammaTild^\alpha_{\mu \nu} = \Gamma^\alpha_{\mu \nu} + 
A_{\mu} \delta^\alpha_\nu ,
\end{eqnarray}
with a general co-vector field $A$ as the independent degree of 
freedom \cite{Dadhich:2012htv}. This field contributes, for instance, to the trace of 
the Cartan torsion, $T_{\mu\alpha}^\alpha=3 A_\mu$ 
\cite{Baekler:2011jt,Shapiro:2001rz}. For the discussion of a relation to 
the affine Weyl connection, see \cite{Sauro:2022chz,Sauro:2022hoh}.

The generalized curvature tensors in the Palatini formulation can now be 
expressed in terms of curvature quantities familiar from ordinary Einstein 
gravity which derive from the Levi-Cevita connection and 
additional terms that depend on the new vector field $A$,
\begin{eqnarray}
\RTild_{\rho \sigma \mu \nu}  &=& R_{\rho \sigma \mu \nu} + 
g_{\rho \sigma} F_{\mu \nu} \label{eq:RTinRF}\\
\RTild_{\sigma \nu} &=& R_{\sigma \nu} + F_{ \sigma \nu} \\
\RTild &=& R, \label{eq:RTildR}
\end{eqnarray}
with the tensor $F$ acquiring the form of a Maxwellian field strength,
\begin{eqnarray}
F_{ \mu \nu} = \partial_\mu A_\nu - \partial_\nu A_\mu .
\end{eqnarray}
In \Eqref{eq:RTildR} we observe that the generalized Ricci scalar reduces to 
the standard Ricci scalar on shell. Therefore, the classical action 
\eqref{eq:EHP} is on-shell equivalent to the Einstein-Hilbert action of 
classical GR, see \cite{Dadhich:2012htv,Janssen:2019htx,G:2022ger} for 
recent detailed discussions. Since each component $A_\mu\in\mathbb{R}$, the 
Einstein-Hilbert-Palatini action has an $\mathbb{R}^4$ gauge invariance. 

Interestingly, the $A$ field appears in the form of a Maxwell-type field 
strength tensor $F$ as the anti-symmetric part of the generalized Ricci tensor. 
At higher orders in the curvature, we can thus expect that more general gravity 
theories of higher order in the curvature will exhibit Maxwellian gauge 
invariance. This implies that the $\mathbb{R}^4$ invariance for this 
part of the general connection reduces to a $U(1)$ invariance at higher orders. 

Let us therefore consider terms to second order in the curvature. More 
specifically, we concentrate on terms that can be constructed from a Ricci-like 
tensor. Since the generalized Riemann tensor $\RTild^{\cdot}{}_{\cdots}$ is not 
anti-symmetric in the first two indices, we can construct a 
second Ricci-like tensor of rank two $\LTild$ by instead 
contracting the second and the fourth index
\begin{eqnarray}
\LTild_{\sigma \nu} &= g^{\rho \mu} \RTild_{\sigma \rho \nu 
\mu} &= R_{\sigma \nu} - F_{ \sigma \nu}.
\end{eqnarray}
The tensors $\LTild$ and $\RTild$ obviously coincide 
in the limit $A\to0$, reducing to the ordinary Ricci tensor $R$. In the general 
case, we can use both curvature tensors for the construction of invariants, 
yielding
\begin{eqnarray}
\RTild_{\sigma \nu} \RTild^{\sigma \nu}= \LTild_{\sigma \nu} \LTild^{\sigma \nu} 
  &= R_{\sigma \nu} 
R^{\sigma \nu} + F_{ \sigma \nu} F^{ \sigma \nu} ,\\
\RTild_{\sigma \nu} \LTild^{\sigma \nu}  &= R_{\sigma \nu} 
R^{\sigma \nu} - F_{ \sigma \nu} F^{ \sigma \nu}.
\end{eqnarray}
We observe that only two combinations are independent. A general contribution 
to the action can thus be spanned by the linear combination of the two 
independent invariants. On shell, we have the equivalence for general couplings 
$\sigma^1, \sigma^2$:
\begin{eqnarray}
& \sigma^1 \RTild_{\mu \nu} \RTild^{\mu \nu} +  \sigma^2 
\RTild_{\mu \nu} \LTild^{\mu \nu} \nonumber \\
=& \ \sigma^R R_{\mu \nu} R^{\mu \nu} + \sigma^F F_{\mu 
\nu} F^{\mu \nu}.
\label{eq:sigma12RF}
\end{eqnarray}
The corresponding couplings in front of the standard Ricci-squared and 
Maxwell terms satisfy
\begin{eqnarray}
\sigma^R &= \sigma^1 + \sigma^2 \\
\sigma^F &= \sigma^1 - \sigma^2.
\label{eq:sigmaRF}
\end{eqnarray}
Equation \eqref{eq:sigma12RF} illustrates that a second order curvature theory 
built from the generalized Ricci-like tensors is on-shell equivalent to a 
second-order metric theory plus an abelian gauge field. Of course, a further 
independent second-order invariant can be formed by suitably squaring the 
generalized Riemann tensor.  As is obvious from \Eqref{eq:RTinRF}, this boils 
down to a square of the Riemann tensor and a Maxwell term as well. In the 
following, we ignore such terms to quadratic order in the Riemann tensor for 
simplicity.

\section{Quantum Hilbert-Palatini gravity}
\label{sec:QEPG}

The preceding observations on the classical level suggest to study the 
quantized version of Hilbert-Palatini gravity in the on-shell reduction scheme: 
we use the degrees of freedom of the on-shell form found for the general 
connection to first order in the curvature, i.e., \Eqref{eq:genPalatiniSol}, to 
quantize the theory to second order in the curvature. In 
practice, this 
corresponds to extending results for quantum Einstein gravity to this order by 
including a Maxwell-type gauge field. 

\subsection{Renormalization flow of Hilbert-Palatini gravity}
\label{subsec:Set-up}

We now investigate the renormalization flow of the gravitational effective 
action $\Gamma_{k}[g, \GammaTild]$ in the theory space spanned by the 
Hilbert-Palatini action including the terms to quadratic order in Ricci-like 
curvature tensors as discussed above, 
\begin{eqnarray}
\Gamma_{\text{gr}, k}[g,\GammaTild] =& \int d^4x \sqrt{g} \frac{1}{16 
\pi 
\bar{G}_k} \left[ \bar{\Lambda}_k - 2 \RTild  + 
 \sigmaBar^1_k \RTild^{\mu \nu} \RTild_{\mu \nu} \right. 
 \nonumber \\ & \left. + \sigmaBar^2_k \LTild^{\mu \nu} 
 \RTild_{\mu \nu}  \right].
\end{eqnarray}
Here, $k$ denotes a renormalization scale at which the theory is considered, 
and all coupling constants are considered to be $k$ dependent. Now, instead of 
considering all degrees of freedom of the general connection $\GammaTild$, we 
perform the on-shell reduction of \Eqref{eq:genPalatiniSol} which allows us to 
understand the action as a functional of the metric and the abelian gauge field,
\begin{eqnarray}
\Gamma_{\text{gr}, k}[g,A] =& \int d^4x \sqrt{g} \frac{1}{16 \pi 
\bar{G}_k} \left[ \bar{\Lambda}_k - 2 R  + 
 \sigmaBar^R_k R^{\mu \nu} R_{\mu \nu} \right. \nonumber \\ 
 & \left. + \sigmaBar^F_k F^{\mu \nu} F_{\mu \nu}  \right] .
\end{eqnarray}
We are interested in the scale dependence of the running 
Newton coupling $\bar{G}_k$, the cosmological parameter $\bar{\Lambda}_k$, the 
higher curvature coupling $\sigmaBar^R_k$, and the wave function 
renormaliztion $Z^A_k$ of the abelian field strength defined by
\begin{eqnarray}
Z^A_k = \frac{\bar{\sigma}^F_k}{4 \pi \bar{G}_k} .
\end{eqnarray}
For a treatment of the gauge degrees of freedom, we use the 
background field formalism and perform a linear split of the 
metric $g$ and the gauge field $A$ into fluctuations around 
their respective background fields which are denoted by a bar
\begin{eqnarray}
g_{\mu \nu} &=& \gBar_{\mu \nu} + \bar{\kappa} h_{\mu \nu}, 
\\
A_{\mu} &=& \ABar_{\mu} + a_{\mu},
\end{eqnarray}
with the abbreviation
\begin{eqnarray}
\bar{\kappa}^2 = 32 \pi \bar{G} .
\end{eqnarray}
The rescaling of the metric fluctuation $h$ by the quantity $\bar{\kappa}$ 
ensures a standard canonical mass dimension of the field.
For the Faddeev-Popov quantization, we include gauge-fixing terms
\begin{eqnarray}
\Gamma_{\text{gf}, k}  = \frac{1}{2} \int d^4 x \sqrt{\gBar} \left( 
\frac{1}{\alpha_{\text{gr}}} \mathcal{F}^\mu \mathcal{F}_\mu + 
\frac{1}{\alpha_{A}} \mathcal{G} \mathcal{G}\right)
\end{eqnarray}
with gauge parameters $\alpha_\text{gr}$ and $\alpha_A$. As 
gauge-fixing conditions for the metric 
sector $\mathcal{F}$ and the abelian gauge sector $
\mathcal{G}$, we use
\begin{eqnarray}
\mathcal{F}^\mu &=& \sqrt{2} \bar{\kappa} \left( \gBar^{\mu 
\kappa} \nablaBar^\lambda - \frac{1+\beta}{4} \gBar^{\kappa 
\lambda} \nablaBar^\mu \right) h_{\kappa \lambda} \\
\mathcal{G} &=& \sqrt{Z^A_k} \left(\nablaBar_\mu a ^\mu 
\right),
\end{eqnarray}
where $\beta$ denotes another gauge parameter of the metric sector. 
Also including the corresponding ghost terms $\Gamma_{\text{gh}, k}$ for both 
sectors, the total effective (average) action reads
\begin{eqnarray}
\Gamma_{k}[\PhiBar, \Phi] = \Gamma_{\text{gr}, k} + \Gamma_{\text{gf}, k} 
+ \Gamma_{\text{gh}, k}.
\end{eqnarray}
Here, $\PhiBar$ and $\Phi$ denote collective field variables, representing the 
background and fluctuations fields, respectively,
\begin{eqnarray}
\left( \Phi \right) &=& \left( h, a, \bar{c}, c, \bar{b}, b \right) \\
\left( \PhiBar \right) &=& \left( \gBar, \ABar \right)
\end{eqnarray}
with $\bar{c}$ and $c$ being the (anti-)ghost fields for the 
gravitational sector and $\bar{b}$ and $b$ being the (anti-)ghost fields for the 
abelian gauge sector.

We quantize the system, using the Wetterich equation 
\cite{Wetterich:1992yh,Ellwanger:1993mw,Morris:1993qb,Bonini:1992vh},
\begin{eqnarray}
\partial_t \Gamma_k[\PhiBar,\Phi] = \frac{1}{2} \text{STr} \left[ 
\left( \Gamma^{(2)}_k[\PhiBar, \Phi] + \mathcal{R}_k 
\right)^{-1}\partial_t \mathcal{R}_k \right],
\label{eq:Wetterich}
\end{eqnarray}
to compute the renormalization flows of the renormalized, 
dimensionless couplings denoted without a bar
\begin{eqnarray}
G_k = k^2 \bar{G}_k, \qquad \Lambda_k = \frac{1}{k^2}
\bar{\Lambda}_k, \qquad \sigma^R_k =  k^2 \sigmaBar^R_k,
\end{eqnarray}
For simplicity, we focus on the Landau gauge, choosing
\begin{eqnarray} 
\alpha_{A} &\rightarrow & 0, \\
\alpha_{\text{gr}} &\rightarrow & 0, \\
\beta &=& 0,
\end{eqnarray}
see \cite{Gies:2015tca,Ohta:2016npm,Ohta:2016jvw,DeBrito:2018hur} for studies 
of gauge or parametrization dependencies in the metric context.
For the computation of the traces and the identification of the corresponding 
operators on both sides, we use a spherical background $\bar{g}$, and a 
covariantly constant background field $\bar{A}$. For the details of the 
regularization around the scale $k$ controlled by the regulator $\mathcal{R}_k$ 
in \Eqref{eq:Wetterich}, we choose a Type I regularization 
scheme, following the computation of \cite{Falls:2017lst}. Computations that 
include further invariants and higher order curvature terms are, in principle, 
possible, e.g., along the lines of 
\cite{Lauscher:2002sq,Benedetti:2009rx,Groh:2011vn,Ohta:2013uca,Gies:2016con,
Falls:2017lst,
Falls:2020qhj,Kluth:2020bdv,Kluth:2022vnq,Sen:2021ffc}

Using the flows for the dimensionless, renormalized couplings, the wave 
function renormalization $Z^A_k$ occurs only through the corresponding
anomalous dimension
\begin{eqnarray}
\eta_A = - \frac{k \partial_k Z^A_k}{Z^A_k},
\end{eqnarray}
which is determined by an algebraic equation. The flows of the couplings as 
driven by the metric fluctuations has been computed in \cite{Falls:2017lst}. 
These are amended by contributions from the abelian vector field which 
we evaluate analogously to \cite{Dona:2013qba}, but to second order in the 
curvature. The anomalous dimension of the abelian gauge field subject to metric 
fluctuations has been computed in \cite{deBrito:2019umw}; see the 
Appendix.

We collect all running couplings into  $\vec{u}$ which is 
a vector in the truncated theory space
\begin{eqnarray}
\vec{u}_k = \begin{pmatrix} G_k\\ \Lambda_k \\ \sigma_k
\end{pmatrix},
\end{eqnarray}
allowing for a compact notation for the flow equations
\begin{eqnarray}
\vec{\beta}\left( \vec{u} \right) = \begin{pmatrix} \beta_G\\ 
\beta_\Lambda \\ \beta_\sigma \end{pmatrix} = 
\begin{pmatrix} k \partial_k G_k \\ k \partial_k \Lambda_k \\ k 
\partial_k \sigma_k \end{pmatrix}.
\end{eqnarray}
The explicit flows are summarized in the Appendix.
We are specifically interested in fixed points $\vec{u}_\star$ of the 
RG flow which satisfy
\begin{eqnarray}
\vec{\beta}\left( \vec{u}_\star \right) \overset{!}{=} 0.
\label{eq:FP}
\end{eqnarray}
In order to characterize the fixed points, we linearize the flow 
equations around the fixed point and determine the critical 
exponents related to the eigenvalues of the Jacobian (stability matrix) of the 
expansion,
\begin{eqnarray}
\{ \theta_1, \theta_2, \theta_3 \} = -\text{eig} \left. \left( 
\vec{\nabla}_{\vec{u}} \otimes \vec{\beta} \right) \right|
_{\vec{u} = \vec{u}_\star} .
\end{eqnarray}
Positive critical exponents characterize RG relevant directions which are 
attracted by the fixed point towards the UV. These directions determine the 
long-range properties of the theory towards the IR and correspond to physical 
parameters.

\subsection{Results}
\label{subsec:Results}

The fixed point equations \eqref{eq:FP} turn out to be rational equations in 
the couplings, see the Appendix, and can be solved analytically. In addition to 
the Gaußian fixed point, we find five non-Gaussian fixed points. Discarding 
those with a negative Newton coupling for physical reasons, those with 
$\Lambda_\ast>\frac{1}{2}$ which is beyond a singularity in the graviton 
propagator, and those with very large values for $G_\ast$ which we consider as 
artifacts of the approximations involved, we end up with one viable fixed 
point the quantitative results of which are listed in Tab.~\ref{tab:FPs}. 

\begin{table}
\begin{tabular}{|c || c | c ||} 
 \hline
{} & Palatini gravity (this work) & Metric gravity \cite{Falls:2017lst}  \\
 \hline\hline
 $G_\star$ & 1.132 & 1.467  \\ 
 \hline
 $\Lambda_\star$ & 0.214 & 0.171  \\
 \hline
 $\sigma_\star$ & 0.326 & 0.339  \\
 \hline \hline
 $\theta_{1,2}$ & 2.057 $\pm$ 3.195 $\cdot \text{i}$ & 1.627 $\pm$ 2.570 $\cdot 
\text{i}$  \\
 \hline
$\theta_{3}$ & 12.780 & 21.232  \\
 \hline
${\eta_{A,\ast}}$ & $-0.0924$ &  --  \\ [1ex] 
 \hline
\end{tabular}
\caption{\label{tab:FPs} Fixed-point solutions and critical exponents in second 
order truncation for Hilbert-Palatini gravity (this work) and metric gravity 
for the present truncation \cite{Falls:2017lst}.}
\end{table}

\begin{figure}[t]
\includegraphics[width=0.48\textwidth]{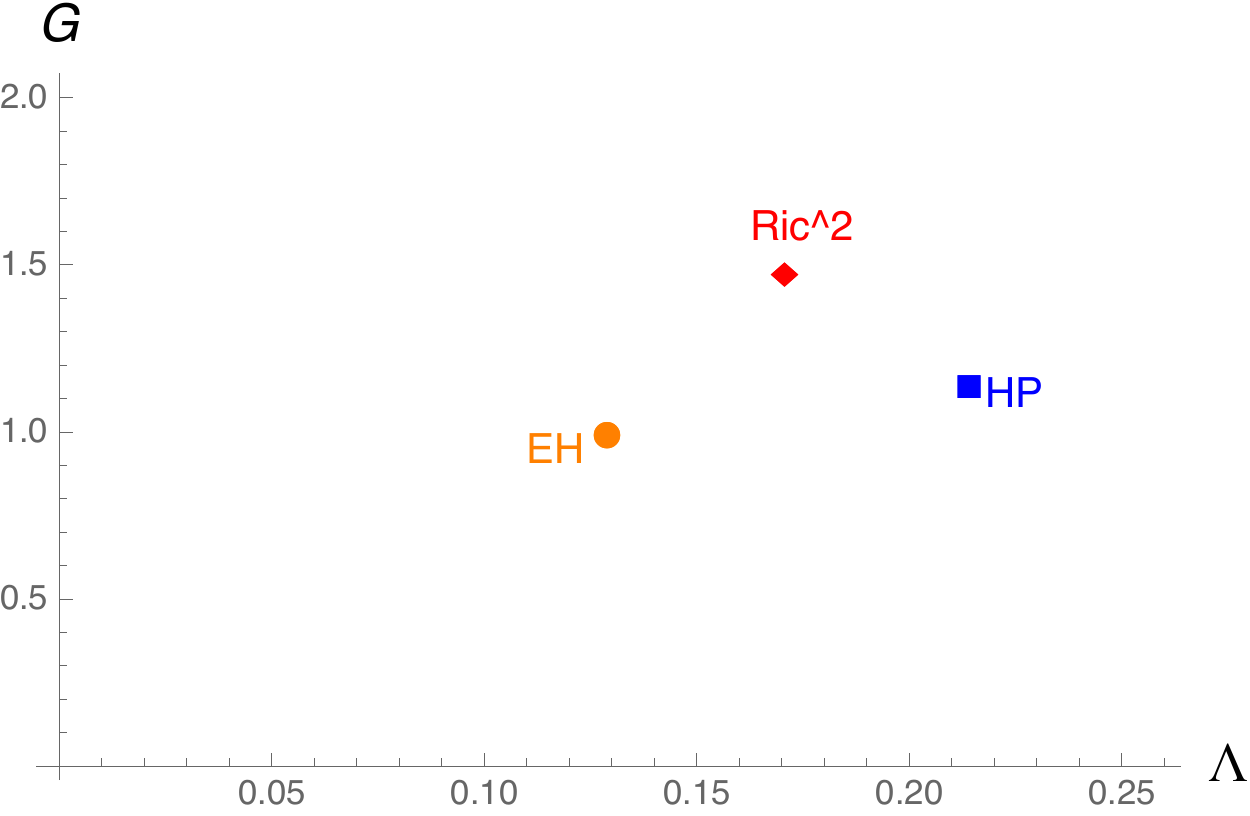}
\caption{Fixed-point positions projected onto the $G,\Lambda$ plane. 
The orange circle and red diamond represent metric gravity in the 
Einstein-Hilbert truncation (EH) and its extension to quadratic order in the 
Ricci 
tensor ($\text{Ric}^2$) \cite{Falls:2017lst}, while the blue square 
represents Hilbert-Palatini gravity (HP) to the second order in the Ricci-
tensor found in this work using on-shell reduction.}
\label{fig:FPpositions}
\end{figure}

For comparison, we also list the results for the Reuter fixed point in pure 
metric gravity obtained in the analogous approximation as obtained in 
\cite{Falls:2017lst}. In general, we observe that the results are rather 
similar to one another which we interpret as evidence that a direct analogue of 
the Reuter fixed point in metric gravity also exists in on-shell reduced 
Hilbert-Palatini gravity with additional dynamical degrees of freedom in the 
general connection. For comparison, we plot the fixed point positions projected 
onto the $G,\Lambda$ plane in Fig.~\ref{fig:FPpositions}. The fixed point 
labeled as ``EH'' marks the Reuter fixed point in metric gravity in the 
lowest-order Einstein-Hilbert truncation. Upon inclusion of terms 
quadratic in the Ricci tensor, this fixed point moves a bit to larger values of 
the coupling 
parameters (labeled by ``Ric${}^2$'' in the figure and listed in the second 
column of Tab.~\ref{tab:FPs}). The position of the corresponding fixed point  
in Hilbert-Palatini found in this work is labeled by ``HP'' in 
Fig.~\ref{fig:FPpositions}.

Inspecting the results of Tab.~\ref{tab:FPs} more closely, we observe that 
specifically the fixed-point value of the Newton coupling is somewhat smaller. 
This can serve as an indication that a quantum gravity theory with independent 
connection variables may more easily be compatible with weak-gravity bounds 
\cite{Eichhorn:2016esv,Eichhorn:2017eht,DeBrito:2019rrh,Kwapisz:2019wrl,
Reichert:2019car, Eichhorn:2020sbo,deBrito:2020dta, 
deBrito:2021pyi,Eichhorn:2019yzm, Laporte:2021kyp, Eichhorn:2021qet} which 
arise from the demand for gravity-matter systems to be compatible with 
particle-physics observations. 

While the leading critical exponents become somewhat larger in Hilbert-Palatini 
gravity, the most decisive change occurs for the third critical exponent 
$\theta_3$ which becomes much smaller by almost a factor of 2. The story of 
this critical exponent is somewhat involved: already in the first analysis of 
the asymptotic safety scenario at the quadratic curvature order 
\cite{Lauscher:2002sq}, this exponents was found to be rather large which 
seemed to contradict the expected hierarchy of decreasing critical exponents 
for higher order operators. In fact, subsequent higher-order truncations 
revealed that this large value $\theta \gtrsim 20$ is a truncation artifact 
\cite{Codello:2007bd,Machado:2007ea,Codello:2008vh}, stabilizing at 
$\mathcal{O}(1)$ if computed at higher order. 
In the light of these findings, we interpret the reduction of $\theta_3$ by a 
factor of 2 as a hint that Hilbert-Palatini gravity may not be so severely 
affected by the truncation artifact. 

It is interesting to observe that the anomalous dimension of the U(1) vector 
field at the fixed point $\eta_{A,\ast}$ is negative. This is in agreement with 
studies of the influence of gravitational fluctuations on (non-)abelian gauge 
fields, where (depending on the matter sector) $\eta_{A,\ast}<0$ can go along 
with either (i) an asymptotically free gauge sector even for abelian gauge 
theories  or (ii) an asymptotically safe gauge sector with a higher degree of 
predictivity \cite{Harst:2011zx,Eichhorn:2017lry,Eichhorn:2017muy}. Both 
scenarios indicate that the fluctuations of the additional degrees of freedom 
in the connection do not induce new UV problems such as Landau pole 
singularities despite their similarity to abelian gauge theories in the 
on-shell reduction scheme.

\begin{figure}[]
\includegraphics[width=0.42\textwidth]{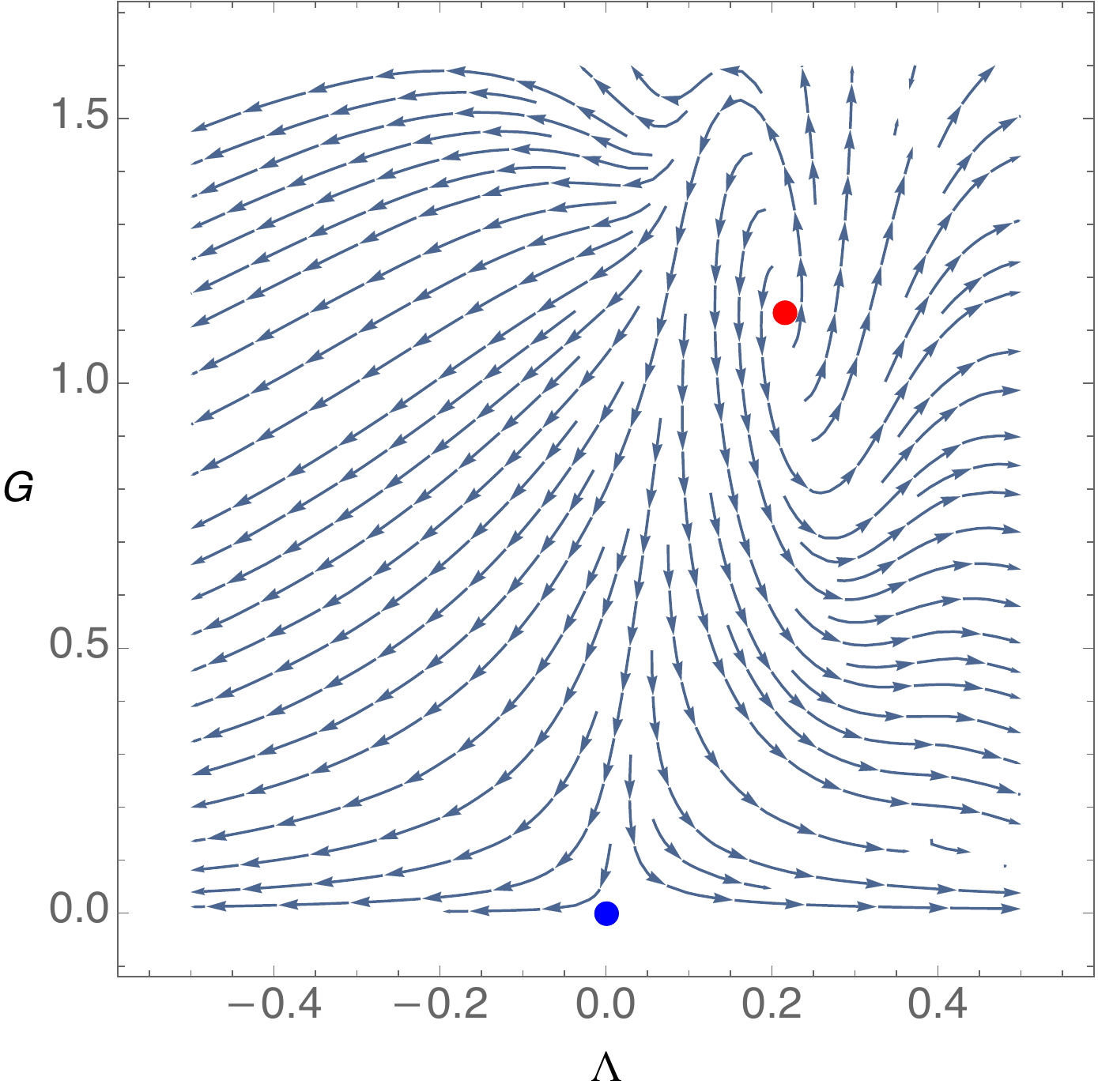}
\caption{Flow diagram in the theory space spanned by 
the couplings $\Lambda$ and $G$ with the third coupling set to 
its UV fixed-point value $\sigma_\star$. The red dot represents 
the non-Gaussian UV fixed-point and the blue dot the Gaussian 
IR fixed-point. The arrows flow towards the IR.
}
\label{fig:streamplot}
\end{figure}

For the physical validity of the fixed point, a crucial question is as to 
whether an RG trajectory exists that connects the high-energy fixed-point 
regime with the regime of classical gravity where the dimensionful renormlized 
Newton coupling and cosmological constant are indeed constant over a wide range 
of scales (higher order curvature couplings are not tightly constraint by 
observations). For this, an RG trajectory must exist that emanates from the UV 
fixed point and passes by sufficiently near the Gaussian fixed point for $G$ 
and $\Lambda$ such that they satisfy canonical scaling. The fact that such 
trajectories exist is illustrated by the stream plot in the $G,\Lambda$ plane 
(evaluated at $\sigma=\sigma_\ast$) in Fig.~\ref{fig:streamplot} 
where the arrows indicate the RG flow towards the IR. We conclude that our 
findings support the existence of a UV-complete RG trajectory in quantum 
Hilbert-Palatini gravity that features a long-range regime where classical GR 
holds as an effective low-energy theory.

\section{Conclusions}
\label{sec:conc}

We have analyzed the renormalization flow of Hilbert-Palatini gravity using the 
functional renormalization group. Our study provides evidence for the existence 
of a non-Gaussian UV fixed point similar to the Reuter-fixed point of metric 
gravity. This result is based on an analysis of an expansion of the action 
in terms of curvature invariants including squares of generalized Ricci-like 
tensors and uses an on-shell reduction scheme that allows to gradually include 
the additional degrees of freedom introduced by a general connection in 
comparison to a pure metric formulation. 

The discovered fixed point supports the existence of UV-complete 
RG trajectories in Hilbert-Palatini gravity within an asymptotic safety scenario. 
Quantitatively, the fixed point occurs at coupling values 
similar to those of metric gravity. A similar comment 
applies to the critical exponents -- although we even find 
indications for a larger degree of stability under the increase of the 
expansion order. Importantly, there exist RG trajectories emanating from this 
fixed point which can be connected to a low-energy regime with the long-range 
limit corresponding to Einstein's classical GR. 

Within our on-shell reduction scheme, the additional degrees of freedom 
in the general connection reduce to a vector field that is related to the trace 
of the Cartan torsion. In our present truncation, the vector field features a 
local U(1) invariance and thus contributes similarly to an abelian gauge 
field. This observation holds true for any truncation built from local 
curvature invariants of the generalized Riemann tensor. 

To higher-orders in the on-shell reduction scheme and the curvature expansion, 
we expect that further additional degrees of freedom acquire their own dynamics 
and start to contribute to the flow. At the present order, they drop out, 
because their equation of motion is algebraic and their action corresponds to 
that of simple quadratic mass terms. Therefore, we expect that their dynamics 
at higher orders corresponds to that of massive modes. This does not only 
suggest that they decouple towards the IR, but are also likely to contribute to 
the UV only at the prize of corresponding mass suppression factors. This 
observation justifies to consider the on-shell reduction scheme as a 
quantitatively controlled expansion scheme, provided the underlying curvature 
expansion scheme has satisfactory convergence properties for observables. 

While the differences to metric gravity as found in this work are comparatively 
small, the question remains as to whether the additional connection degrees of 
freedom exert a stronger influence on other sectors. In particular, since the 
additional field after on-shell reduction is a vector field resembling an 
abelian gauge field, a possible impact on the matter sector in the fixed-point 
regime or beyond is conceivable. Towards low-energies, this degree of freedom 
has been discussed as a candidate for dark matter (dark photon) 
\cite{Pradisi:2022nmh}. A particularly relevant question for the high-energy 
regime where gravity is non-perturbative refers to possible consequences of 
this degree of freedom for the realization of symmetries such as chiral 
symmetry of fermionic matter particles 
\cite{Eichhorn:2011pc,Meibohm:2016mkp,Eichhorn:2016vvy,Eichhorn:2018nda,
Gies:2018jnv,Gies:2021upb}. The latter is closely related to the existence of 
light fermions in nature which are an observational fact that needs to be 
supported also by the quantum gravitational sector. 

Finally, we believe that our on-shell reduction scheme can also be useful 
in further formulations of quantum gravity with different and/or additional 
degrees of freedom. An example would be given by ``tetrad-only'' formulations 
\cite{Harst:2012ni} or generalizations of Hilbert-Palatini gravity using the 
spin-base invariant formalism \cite{Gies:2013noa}. In the latter case, it has 
been shown that on-shell reduction of a generalized spin connection would 
entail two  
vector fields \cite{Kuespert:2020}, presumably with analogous consequences for 
the construction of an asymptotic safety scenario as found in the present work. 

Beyond quantum gravity, on-shell reduction is a rather obvious scheme in 
functional RG approaches to supersymmetric theories 
\cite{Synatschke:2008pv}. In this case, on-shell reduction 
eliminates the auxiliary field(s) that are introduced for a superfield 
formulation in superspace. While the functional RG can be employed both in the 
on-shell as well 
as the off-shell case, the off-shell formulation is advantageous for the 
description of phase transitions in connection with order parameters related to 
the off-shell sector 
\cite{Gies:2009az,Synatschke:2009nm,Heilmann:2012yf,Gies:2017tod}. As a word of 
caution, it may therefore be advisable not to use the on-shell reduction scheme 
for systems in which the off-shell sector is relevant for critical phenomena.

\section*{Acknowledgments}

We thank Astrid Eichhorn, Aaron Held, Yannick Kluth, Benjamin 
Knorr, Ruben Küspert, Oleg Melichev, Gustavo Pazzini de Brito, 
Roberto Percacci, Marc Schiffer, and Kevin Tam for 
valuable discussions. ASS acknowledges the hospitality of CP3-Origins at Odense 
where part of this work was done. 
This work has been funded by the Deutsche 
Forschungsgemeinschaft (DFG) under Grant No. 406116891 
within the Research Training Group RTG 2522/1.
\\
\\
\appendix

\section{Flow equations}

The anomalous dimension for an abelian gauge field $\eta^{A}$ has been 
computed in Appendix D of Ref.~\cite{deBrito:2019umw} and expressed in 
our notation reads
\begin{eqnarray}
\eta^{A} = -\frac{G (10-40 \Lambda +7 \sigma )}{18 \pi  (1-2 \Lambda +
\sigma )^2}.
\end{eqnarray}

The flow equations for the three remaining couplings $\Lambda$, $G$ 
and $\sigma$ have been computed along the lines of \cite{Falls:2017lst}. The 
right-hand side of the Wetterich equation -- labeled as $I$ in 
\cite{Falls:2017lst} -- is extended by the contributions from the abelian gauge 
field 
to quadratic order in the curvature according to \cite{Percacci:2017fkn}. 
The final results for the flow equations are
\begin{eqnarray}
\beta_{\Lambda} = -\frac{A_\Lambda}{B_\Lambda}, \qquad \beta_{G} = \frac{A_G}{B_G}, \qquad \beta_{\sigma} = 2\frac{A_\sigma}{B_\sigma},
\end{eqnarray}
with
\begin{widetext}

\begin{eqnarray}
A_{\Lambda} = &-432 \pi ^2 G^2 (-2 \Lambda +\sigma +1)^2 \cdot A^{(2)}_\Lambda  -67184640 \pi ^4 \Lambda  (4 \Lambda +6 \sigma -3)^3 (-2 \Lambda +\sigma +1)^5 + G^4 \cdot A^{(4)}_{\Lambda} \nonumber \\
&+ 31104 \pi ^3 G (-2 \Lambda +\sigma +1)^2 \cdot A^{(1)}_{\Lambda} + 3 \pi  G^3 \cdot A^{(3)}_{\Lambda}
\end{eqnarray}

\begin{eqnarray}
A^{(1)}_{\Lambda} =&-27 \Lambda  \left(16200 \sigma ^6+90996 \sigma ^5-92998 \sigma ^4-145681 \sigma ^3+120465 \sigma ^2-8599 \sigma -5701\right) \nonumber \\
& +622080 \Lambda ^7+128 \Lambda ^6 (9000 \sigma -15743)-32 \Lambda ^5 \left(71460 \sigma ^2+48998 \sigma -87891\right) \nonumber \\
&+18 \Lambda ^2 \left(79200 \sigma ^5-188908 \sigma ^4-502556 \sigma ^3+553149 \sigma ^2-80446 \sigma -25643\right)  \nonumber \\
&+12 \Lambda ^3 \left(101880 \sigma ^4+825560 \sigma ^3-1166136 \sigma ^2+173151 \sigma +96557\right) \nonumber \\
&-8 \Lambda ^4 \left(483840 \sigma ^3-1212552 \sigma ^2+53236 \sigma +279421\right) \nonumber \\
&+7290 \left(2 \sigma ^2+\sigma -1\right)^2 \left(20 \sigma ^2+\sigma -4\right) 
\end{eqnarray}
\begin{eqnarray}
A^{(2)}_{\Lambda} =& 5633536 \Lambda ^6+64 \Lambda ^5 (463464 \sigma -478897)+32 \Lambda ^4 \left(865786 \sigma ^2-3584586 \sigma +1918663\right) \nonumber \\
&+ 27 \Lambda  \left(79720 \sigma ^5+712632 \sigma ^4-2374666 \sigma ^3-1147514 \sigma ^2+1342848 \sigma -200281\right) \nonumber \\
&+ 36 \Lambda ^2 \left(160900 \sigma ^4+1825044 \sigma ^3+1414833 \sigma ^2-3232750 \sigma +848251\right) \nonumber \\
&- 243 \left(41500 \sigma ^5-43496 \sigma ^4+14055 \sigma ^3-31589 \sigma ^2+5968 \sigma +1608\right) \nonumber \\
&- 12 \Lambda ^3 \left(2566696 \sigma ^3+5505748 \sigma ^2-13621200 \sigma +5113385\right)
\end{eqnarray}
\begin{eqnarray}
A^{(3)}_{\Lambda} =&-54 \Lambda  \left(242816 \sigma ^6-43621136 \sigma ^5-105741110 \sigma ^4-33856598 \sigma ^3+65125975 \sigma ^2-58567640 \sigma +19071545\right) \nonumber \\
&-81 \left(3694416 \sigma ^6+14436454 \sigma ^5+16191029 \sigma ^4-889985 \sigma ^3-4324109 \sigma ^2+5270879 \sigma -1603988\right)  \nonumber \\
&+36 \Lambda ^2 \left(3194560 \sigma ^5-152974904 \sigma ^4-131239628 \sigma ^3+194163574 \sigma ^2-197711957 \sigma +83407825\right) \nonumber \\
& 106823680 \Lambda ^7+512 \Lambda ^6 (387008 \sigma -938935)-64 \Lambda ^5 \left(661408 \sigma ^2+23286680 \sigma -1885541\right) \nonumber \\
&-24 \Lambda ^3 \left(12370624 \sigma ^4-146478720 \sigma ^3+172896708 \sigma ^2-225254630 \sigma +165900405\right) \nonumber \\
&+16 \Lambda ^4 \left(24289856 \sigma ^3+27106560 \sigma ^2+18900738 \sigma +133505315\right) 
\end{eqnarray}
\begin{eqnarray}
A^{(4)}_{\Lambda} =& 67230720 \Lambda ^5+128 \Lambda ^4 (578163 \sigma -1781875)-16 \Lambda ^3 \left(2661552 \sigma ^2+8298109 \sigma -18340280\right) \nonumber \\
&+27 \left(1232336 \sigma ^5+5064046 \sigma ^4+5332307 \sigma ^3+620285 \sigma ^2-500522 \sigma -193960\right) \nonumber \\
&-18 \Lambda  \left(15761360 \sigma ^4+36304598 \sigma ^3+6154946 \sigma ^2-3194187 \sigma -2801570\right)\nonumber \\
&+12 \Lambda ^2 \left(44960536 \sigma ^3+7553626 \sigma ^2+425125 \sigma -14839910\right) 
\end{eqnarray}
\\
\begin{eqnarray}
B_\Lambda = 216 \pi ^2 (-2 \Lambda +\sigma +1)^2 \cdot \left[ 155520 \pi ^2 \left(8 \Lambda ^2+8 \Lambda  \sigma -10 \Lambda -6 \sigma ^2-3 \sigma +3\right)^3 + G^2\cdot B^{(2)}_{\Lambda} + 144 \pi  G \cdot B^{(1)}_{\Lambda}\right]
\end{eqnarray}

\begin{eqnarray}
B^{(1)}_{\Lambda}=& 101248 \Lambda ^5+64 \Lambda ^4 (4693 \sigma -7716)+8 \Lambda ^3 \left(27216 \sigma ^2-180968 \sigma +119287\right)\nonumber \\
&+27 \left(3012 \sigma ^5+9284 \sigma ^4+11483 \sigma ^3-26775 \sigma ^2+15752 \sigma -2887\right) \nonumber \\
&-18 \Lambda  \left(14444 \sigma ^4+17728 \sigma ^3-99747 \sigma ^2+94133 \sigma -23561\right) \nonumber \\ 
&+12 \Lambda ^2 \left(12640 \sigma ^3-106428 \sigma ^2+201663 \sigma -75629\right) 
\end{eqnarray}
\begin{eqnarray}
B^{(2)}_{\Lambda}=& 334208 \Lambda ^4+32 \Lambda ^3 (8839 \sigma -14089)+48 \Lambda ^2 \left(487 \sigma ^2+49392 \sigma -11720\right) \nonumber \\
&-27 \left(51700 \sigma ^4+197344 \sigma ^3+129897 \sigma ^2-108638 \sigma +13984\right) \nonumber \\
&+18 \Lambda  \left(230868 \sigma ^3+155580 \sigma ^2-325499 \sigma +57664\right) 
\end{eqnarray}
\\
\\
\begin{eqnarray}
A_G =& 2 G (4 \Lambda +6 \sigma -3)^3 \left[ -3 \cdot A^{(1)}_{G} \cdot A^{(2)}_{G}  + G^2 \cdot A^{(3)}_{G}  \cdot A^{(4)}_{G} \right]
\end{eqnarray}

\begin{eqnarray}
A^{(1)}_{G} =& 30 \pi  G (4 \Lambda +6 \sigma -3)^2 \left(-6 \Lambda  (5 \sigma +3)+9 \sigma ^2+44 \sigma +25\right) \nonumber \\
& -54 \pi  G \left(4 \Lambda  (4 \sigma -3)+72 \sigma ^2-46 \sigma +9\right) (-2 \Lambda +\sigma +1)^2 \nonumber \\
&+243 \pi  G (4 \Lambda +6 \sigma -3)^2 (-2 \Lambda +\sigma +1)^2 \nonumber \\
&-432 \pi ^2 (4 \Lambda +6 \sigma -3)^2 (-2 \Lambda +\sigma +1)^2 \nonumber \\
&G^2 (4 \Lambda +6 \sigma -3)^2 (40 \Lambda -7 \sigma -10)
\end{eqnarray}
\begin{eqnarray}
A^{(2)}_{G} =& -6 G \left(496 \Lambda ^2+24 \Lambda  (22 \sigma -31)-9 \left(276 \sigma ^2+44 \sigma -31\right)\right) (-2 \Lambda +\sigma +1)^3 \nonumber \\
&+5 G (4 \Lambda +6 \sigma -3)^3 \left(116 \Lambda ^2-4 \Lambda  (19 \sigma +129)+24 \sigma ^2+118 \sigma +469\right) \nonumber \\
&+1080 \pi  (-4 \Lambda -6 \sigma +3)^3 (-2 \Lambda +\sigma +1)^3 
\end{eqnarray}
\begin{eqnarray}
A^{(3)}_{G} =& 24 (4 \Lambda +15 \sigma -3) (-2 \Lambda +\sigma +1)^2+5 (4 \Lambda +6 \sigma -3)^2 (20 \Lambda -7 \sigma -22) 
\end{eqnarray}
\begin{eqnarray}
A^{(4)}_{G} =& 54 \pi  \left(16 \Lambda ^2 (62 \sigma +29)+24 \Lambda  \left(4 \sigma ^2+86 \sigma -29\right)+9 \left(-1192 \sigma ^3+788 \sigma ^2-234 \sigma +29\right)\right) (-2 \Lambda +\sigma +1)^3 \nonumber \\
&+90 \pi  (-4 \Lambda -6 \sigma +3)^3 \left(4 \Lambda ^2 (29 \sigma -1)-2 \Lambda  \left(43 \sigma ^2+351 \sigma +178\right)+44 \sigma ^3+117 \sigma ^2+822 \sigma +499\right)  \nonumber \\
& -G (-4 \Lambda -6 \sigma +3)^3 (-40 \Lambda +7 \sigma +10) (-2 \Lambda +\sigma +1) \nonumber \\
&+7038 \pi  (-4 \Lambda -6 \sigma +3)^3 (-2 \Lambda +\sigma +1)^3  
\end{eqnarray}
\\
\begin{eqnarray}
B_G =& 9 \pi  (-4 \Lambda -6 \sigma +3)^5 (-2 \Lambda +\sigma +1)^2 \left[ -155520 \pi ^2 \left(8 \Lambda ^2+8 \Lambda  \sigma -10 \Lambda -6 \sigma ^2-3 \sigma +3\right)^3 \right. \nonumber \\
& \left. + G^2 \cdot B^{(2)}_{G} +144 \pi  G \cdot B^{(1)}_{G} \right]
\end{eqnarray}

\begin{eqnarray}
B^{(1)}_{G} =& 101248 \Lambda ^5+64 \Lambda ^4 (4693 \sigma -7716)+8 \Lambda ^3 \left(27216 \sigma ^2-180968 \sigma +119287\right) \nonumber \\
&+27 \left(3012 \sigma ^5+9284 \sigma ^4+11483 \sigma ^3-26775 \sigma ^2+15752 \sigma -2887\right) \nonumber \\
&-18 \Lambda  \left(14444 \sigma ^4+17728 \sigma ^3-99747 \sigma ^2+94133 \sigma -23561\right) \nonumber \\
&+12 \Lambda ^2 \left(12640 \sigma ^3-106428 \sigma ^2+201663 \sigma -75629\right) 
\end{eqnarray}
\begin{eqnarray}
B^{(2)}_{G} =& -334208 \Lambda ^4-32 \Lambda ^3 (8839 \sigma -14089)-48 \Lambda ^2 \left(487 \sigma ^2+49392 \sigma -11720\right) \nonumber \\
&+27 \left(51700 \sigma ^4+197344 \sigma ^3+129897 \sigma ^2-108638 \sigma +13984\right) \nonumber \\
&-18 \Lambda  \left(230868 \sigma ^3+155580 \sigma ^2-325499 \sigma +57664\right)
\end{eqnarray}
\\
\\
\begin{eqnarray}
A_{\sigma} =& -1399680 \pi ^3 \sigma  (-2 \Lambda +\sigma +1)^5 (4 \Lambda +6 \sigma -3)^3 + 9 \pi  G^2 (-2 \Lambda +\sigma +1)^2 \cdot A^{(2)}_{\sigma} \nonumber \\
&-648 \pi ^2 G (-2 \Lambda +\sigma +1)^2 \cdot A^{(1)}_{\sigma} +G^3 \cdot A^{(3)}_{\sigma}
\end{eqnarray}

\begin{eqnarray}
A^{(1)}_{\sigma} =&-27 \left(16200 \sigma ^7+32644 \sigma ^6-41054 \sigma ^5-133133 \sigma ^4+105227 \sigma ^3+29329 \sigma ^2-41471 \sigma +8812\right)  \nonumber \\
&+18 \Lambda  \left(79200 \sigma ^6-182460 \sigma ^5-431284 \sigma ^4+633841 \sigma ^3+11688 \sigma ^2-255583 \sigma +74494\right) \nonumber \\
&+12 \Lambda ^2 \left(101880 \sigma ^5+826616 \sigma ^4-1214152 \sigma ^3+27963 \sigma ^2+743483 \sigma -278368\right) \nonumber \\
&-8 \Lambda ^3 \left(483840 \sigma ^4-1152328 \sigma ^3-432924 \sigma ^2+1510489 \sigma -629598\right) \nonumber \\
& 512 \Lambda ^6 (1215 \sigma -1564)+128 \Lambda ^5 \left(9000 \sigma ^2-37931 \sigma +23614\right) \nonumber \\
&-32 \Lambda ^4 \left(71460 \sigma ^3+126606 \sigma ^2-340887 \sigma +156880\right)
\end{eqnarray}
\begin{eqnarray}
A^{(2)}_{\sigma}=& 27 \left(79720 \sigma ^6+174652 \sigma ^5-1081722 \sigma ^4+689497 \sigma ^3+249706 \sigma ^2-327888 \sigma +72432\right) \nonumber \\
&+18 \Lambda  \left(321800 \sigma ^5+1596732 \sigma ^4-2697526 \sigma ^3+781767 \sigma ^2+915448 \sigma -386209\right) \nonumber \\
&-24 \Lambda ^2 \left(1283348 \sigma ^4+242426 \sigma ^3+987572 \sigma ^2-319735 \sigma -233846\right) \nonumber \\ 
& 128 \Lambda ^5 (44012 \sigma +27247)+128 \Lambda ^4 \left(231732 \sigma ^2+1394 \sigma -75545\right) \nonumber \\
&+8 \Lambda ^3 \left(3463144 \sigma ^3-3908896 \sigma ^2-2731276 \sigma +691313\right) 
\end{eqnarray}
\begin{eqnarray}
A^{(3)}_{\sigma}=& 27 \left(30352 \sigma ^7+131896 \sigma ^6+424260 \sigma ^5+105248 \sigma ^4-509251 \sigma ^3+49896 \sigma ^2+147724 \sigma -42080\right) \nonumber \\
&-18 \Lambda  \left(399320 \sigma ^6+1353488 \sigma ^5+1610334 \sigma ^4-4440304 \sigma ^3+753902 \sigma ^2+1625577 \sigma -561570\right)\nonumber \\
&+24 \Lambda ^2 \left(773164 \sigma ^5+1232404 \sigma ^4-6465153 \sigma ^3+3676846 \sigma ^2+2410429 \sigma -1353600\right)\nonumber \\ 
&-8 \Lambda ^3 \left(3036232 \sigma ^4-10230572 \sigma ^3+19821964 \sigma ^2+2913489 \sigma -6181090\right)\nonumber \\
& -5120 \Lambda ^6 (1304 \sigma +189)-128 \Lambda ^5 \left(96752 \sigma ^2-253283 \sigma -94070\right) \nonumber \\
&+64 \Lambda ^4 \left(41338 \sigma ^3+1428338 \sigma ^2-555863 \sigma -578655\right)
\end{eqnarray}
\\
\begin{eqnarray}
B_{\sigma}=& 9 \pi  (-2 \Lambda +\sigma +1)^2 \left[ 155520 \pi ^2 \left(8 \Lambda ^2+8 \Lambda  \sigma -10 \Lambda -6 \sigma ^2-3 \sigma +3\right)^3+ G^2 \cdot B^{(2)}_{\sigma} + 144 \pi  G \cdot B^{(1)}_{\sigma} \right]
\end{eqnarray}

\begin{eqnarray}
B^{(1)}_{\sigma}=& 101248 \Lambda ^5+64 \Lambda ^4 (4693 \sigma -7716)+8 \Lambda ^3 \left(27216 \sigma ^2-180968 \sigma +119287\right) \nonumber \\
&+27 \left(3012 \sigma ^5+9284 \sigma ^4+11483 \sigma ^3-26775 \sigma ^2+15752 \sigma -2887\right) \nonumber \\
&-18 \Lambda  \left(14444 \sigma ^4+17728 \sigma ^3-99747 \sigma ^2+94133 \sigma -23561\right) \nonumber \\
&+12 \Lambda ^2 \left(12640 \sigma ^3-106428 \sigma ^2+201663 \sigma -75629\right) 
\end{eqnarray}
\begin{eqnarray}
B^{(2)}_{\sigma}=& 334208 \Lambda ^4+32 \Lambda ^3 (8839 \sigma -14089)+48 \Lambda ^2 \left(487 \sigma ^2+49392 \sigma -11720\right) \nonumber \\
&-27 \left(51700 \sigma ^4+197344 \sigma ^3+129897 \sigma ^2-108638 \sigma +13984\right) \nonumber \\
&+18 \Lambda  \left(230868 \sigma ^3+155580 \sigma ^2-325499 \sigma +57664\right)
\end{eqnarray}

\end{widetext}

\bibliography{bibliography}

\end{document}